\documentclass[11pt,letterpaper]{article}
\def\TITLE{How not to factor a miracle}

\usepackage[table]{xcolor}
\usepackage{latexsym,graphics,epsfig,multicol}
\usepackage{amsfonts,amssymb,amsmath}

\usepackage{wasysym}

\usepackage{tikz}
\usetikzlibrary{mindmap,trees,backgrounds}

\definecolor{darkgreen}{rgb}{0.0625,0.64,0.0625}
\definecolor{darkblue}{rgb}{0.0625,0.0625,.5}
\definecolor{darkred}{rgb}{.7,.2,.2}
\definecolor{verylightgray}{rgb}{.95,.95,.95}

\usepackage[
pdftitle={\TITLE},  
pdfauthor={Derek K. Wise},
pdfdisplaydoctitle,
colorlinks,%
linkcolor=black,citecolor=red,urlcolor=black,
urlbordercolor=white
]{hyperref}

\usepackage{fancyhdr}
    \makeatletter
    \renewcommand\section{\@startsection{section}{1}{\z@}%
                                      {-3.5ex \@plus -1ex \@minus -.2ex}%
                                      {2.3ex \@plus.2ex}%
                                      {\normalfont\large\bfseries\sffamily\raggedright}}
    \renewcommand\subsection{\@startsection{subsection}{1}{\z@}%
                                      {-3.5ex \@plus -1ex \@minus -.2ex}%
                                      {2.3ex \@plus.2ex}%
                                      {\normalfont\normalsize\bfseries\sffamily}}

    \makeatother

\newcommand{\arxiv}[1]{\href{http://arxiv.org/abs/#1}{\nolinkurl{#1}}}

\begin{document}

\begin{center}
 {\LARGE \sf \TITLE\footnote{Third Prize in the 2015  FQXi Essay Contest, Trick or Truth: the Mysterious Connection Between Physics and Mathematics.  This revised version, 15 December 2015.}}\\[.5em]{\sf Derek K.\ Wise} 
\end{center}
 \vskip 4em
\noindent 
Mathematics is a bit like Zen, in that its greatest masters are likely to deny there being any succinct expression of what it is.  It may seem ironic that the one subject which demands absolute precision in its definitions would itself defy definition, but the truth is, we are still figuring out what mathematics is.  And the only real way to figure this out is to {\em do} mathematics.

Mastering any subject takes years of dedication, but mathematics takes this a step further: it takes years before one even begins to see what it is that one has spent so long mastering.   I say ``begins to see'' because so far I have no reason to suspect this process terminates.  Neither do wiser and more experienced mathematicians I have talked to.

In this spirit, for example, \href{http://press.princeton.edu/titles/8350.html}{\sl The Princeton Companion to Mathematics} \cite{Gowers}, expressly renounces any tidy answer to the question ``What is mathematics?"  Instead, the book replies to this question with 1000 pages of expositions of topics {\em within} mathematics, all written by top experts in their own subfields.  This is a wise approach: a shorter answer would be not just incomplete, but necessarily misleading.

Unfortunately, while mathematicians are often reluctant to define mathematics, others are not.  In 1960, despite having made his own mathematically significant contributions, physicist \href{http://www-history.mcs.st-andrews.ac.uk/Biographies/Wigner.html}{Eugene Wigner} defined mathematics as ``the science of skillful operations with concepts and rules invented just for this purpose" \cite{Wigner}.   This rather negative characterization of mathematics may have been partly tongue-in-cheek, 
but he took it seriously enough to build upon it an argument that mathematics is ``unreasonably effective'' in the natural sciences---an argument which has been unreasonably influential among scientists ever since.

What weight we attach to Wigner's claim, and the view of mathematics it promotes, has both metaphysical and practical implications for the progress of mathematics and physics.   
If the effectiveness of mathematics in physics is a `miracle,' then this miracle may well run out. In this case, we are justified in keeping the two subjects `separate' and hoping our luck continues.  If, on the other hand, they are deeply and rationally related, then this surely has consequences for how we should do research at the interface.  

In fact, I shall argue that what has so far been unreasonably effective is not mathematics but {\em reductionism}---the practice of inferring behavior of a complex problem by isolating and solving manageable `subproblems'---and that physics may be reaching the limits of effectiveness of the reductionist approach.  In this case, {\em mathematics} will remain our best hope for progress in physics, {by finding precise ways to go beyond reductionist tactics}. 

\section*{Is physics unreasonably well described in the language invented for that purpose?}

The essence of Wigner's claim, based an unsupported characterization of mathematics as an elaborate mental game that has nothing {\em a priori} to do with physics, is that mathematics is somehow nonetheless amazingly effective in describing physics.  Were physics and mathematics disjoint endeavors, one empirical, one cognitive and largely arbitrary, then it would indeed be astounding, even `miraculous,' to find them interacting as fruitfully and precisely as they do.  But this viewpoint disregards both history and the mathematical biases set up by our place in the natural world. 

First, historically, the distinction between mathematics and physics is relatively recent, with the movement toward `pure' mathematics progressing gradually since the 18th century.  The historical counterargument to Wigner's thesis has been taken up by mathematics historians \cite{Grattan-Guinness} in detail, and that is not my purpose here.  However, it is worth pointing out that, for example, calculus and a large part of differential equations---still among the physicist's most important tools---were designed precisely for physical applications.  It is at least anachronistic, but bordering on absurd, to claim these are unduly effective in the very subject they owe their existence to.  

In fact, the joint development of physics and mathematics predates the recognition of physics as a separate science.  I must agree with Einstein that ``mathematics generally, and particularly geometry, owes its existence to the need which was felt of learning something about the relations of real things to one another,''  and that ``we may in fact regard [geometry] as the most ancient branch of physics''  \cite{Einstein2}.

Second, at a deep level, mathematics---or rather our limited view of it---is founded on our perception of the natural world.  To get an idea of how much our view of mathematics depends on physics, we need not imagine a world with drastically different physical laws: we can simply consider our own world at different scales, where subtle physical phenomena become readily apparent.  

If we were very {\em large} mathematicians, or equivalently if Newton's gravitational constant $G$ were large, Einstein's famous curving of space and time would be noticeable on the scale of everyday life.  With no rigid spatial backdrop, and no independence of `space' and `time' on ordinary scales, we would at least have a very different history of {geometry}. 

Mathematics would be much more radically different if instead we were very {\em small} mathematicians, or equivalently if Planck's constant $\hbar$, which regulates the level of `quantum fuzziness,' were large.  This would also have affected our ideas of geometry, since there would be no stationary objects with well-defined positions to serve as reference points for measurements.  But also,  we may well have developed little interest in numbers or counting, since distinct collections of objects become much less sensible at quantum scales.  We would thus have less motivation to invent number theory, set theory, or combinatorics.   These subjects, fundamental as they are to mathematics as we know it, enjoy this status because of our place within the natural world.  

Moreover, as `quantum mathematicians' we would presumably prove theorems not according to the familiar rules of logic, which are deeply tied to set theory, but according to rules of \href{http://en.wikipedia.org/wiki/Quantum_logic}{quantum logic}, which reflect the fuzzy, indistinct nature of propositions about fundamentally quantum mechanical systems. Likewise, our foundational notions of probability and statistics would necessarily be quantum in nature.

More radical still, we can consider a scenario in which $G$ and $\hbar$ are both large, corresponding to a realm of experience where effects of `quantum gravity' should be evident.  Here the foundations of mathematics would surely differ even more radically from our usual ones, in ways that one can only speculate on depending on which approach to quantum gravity, if any, one believes.

While it is interesting to consider how our foundations of mathematics are determined by physics, it is in fact recognized {\em from a perspective entirely within pure mathematics} that mathematics itself could be built on alternative foundations.  This goes beyond the philosophy of mathematics to what we might call the {\em mathematics of mathematics}, in that the types of foundations of mathematics---different `mathematical worlds' with their own \href{http://ncatlab.org/nlab/show/internal+logic}{notions of logic}---can be classified mathematically.    For example, the potential `mathematical foundations' with many of the same formal properties as the usual set-based foundations are classified by \href{http://math.ucr.edu/home/baez/topos.html}{topos theory}.  This motivates attempts to use topos theory in the foundations of physics \cite{Isham}.  

If our experience of the physical world were different, then we would have invented very different mathematics, but it would still be mathematics.  It would still be quite effective at describing the world it was built to represent---presumably even more effective in treating certain phenomena.

\section*{Appropriate abstraction} 

Of course, one might object that while some mathematics has been invented specifically for physical applications, it remains surprising that `{\em abstract}' mathematics seems to keep finding its way back into physics.  In fact, this confusion arises when we erroneously conflate the precise definition of `abstract' with its colloquial one.   

Outside of mathematical or philosophical discussions, when people call a thing `abstract,' they may only mean it is difficult to understand, or even unnecessarily cerebral.  But true \href{http://en.wikipedia.org/wiki/Abstraction_%28mathematics%29}{abstraction} 
is just the opposite: a path to {\em clarity}.  
It means stripping away all inessential details, arriving at an idealization whose life purpose is to exemplify the properties or concepts under consideration, and nothing else.  
Russell used the word in this sense when he said the language of physics is necessarily abstract, in order to ``say as little as the physicist means to say.''

This raises a key question: How does one arrive at a {\em good} mathematical abstraction? It is perhaps not surprising that we can find useful mathematical abstractions of the {\em simplest} objects in our experience; ultimately, our ability to find good mathematical abstractions of concepts in physics, especially as compared to other sciences, must rest on physics being much simpler than other sciences.

Much more surprising---and I gather this is the part Wigner also finds incredible---is the ability to `stay' at an abstract mathematical level, not referring back to direct physical experience, and still somehow manage to arrive at new mathematical constructs which later turn out to have applications to physics.  

So, let us ask: Where do new mathematical concepts come from?  
Sadly, Wigner is again rather dismissive on this point, claiming  that most advanced mathematical concepts are defined just so that the mathematician can ``demonstrate his ingenuity and sense of formal beauty." \cite{Wigner}

In fact, defining new concepts in mathematics is a delicate art, and evidently one of the hardest to master.  Whereas a decent Ph.D.\ student in mathematics will 
be able to prove some quite difficult theorems---this is the most obvious skill that a graduate education in mathematics teaches---there is typically no training on what makes a good definition.  Finding the right definitions is simply too advanced. Some definitions, once one recognizes a need for them, are rather obvious.  But in many cases, it takes years for a community of mathematicians to agree on the best definition for a concept.  

To understand what makes a good mathematical definition, it is of course best to study many examples.  Here I will consider just one, but deliberately choose an example that does not yet have well-established physical applications.  One of my current favorite examples is the notion of {\em quantum groupoid}, also called a Hopf algebroid \cite{Boehm,DayStreet}.  \href{http://en.wikipedia.org/wiki/Groupoid}{Groupoids} are a modern mathematical way to study {\em symmetry}---arguably the most important concept in mathematical physics.  Intuitively, quantum groupoids result from taking the concept of groupoid, based on sets and hence classical logic, and importing it into the world of vector spaces and hence quantum logic.  

But realizing this intuitive idea and settling on a definition of quantum groupoid took hard work.  The definition that finally emerged looks quite complicated and, frankly, esoteric from the perspective of an outsider showing up on the scene after the dust has settled.  It is much more complicated than what I would instinctively try, or what researchers did in fact try first.  If you handed me the definition of quantum groupoid out of the blue, I would think it very unlikely to be useful in physics.

So, what makes this definition `good' or `correct' in some sense?   Consider this evidence: 
\begin{itemize}
\itemsep 0em
\item The definition naturally relates mathematical structures with many known applications, including `groupoids' and `quantum groups.' 
\item More naive attempts at the definition were refined by the need to include natural examples.  
\item Despite being rather complicated, equivalent definitions were discovered {\em independently} by researchers with quite different starting points, from within different subfields of mathematics. 
\end{itemize} 

Notice that all of this evidence for the definition's quality (and more could be listed) is purely mathematical, and none of it is based on our desire to ``demonstrate [our] ingenuity and sense of formal beauty.''  On the contrary, the evidence gives a strong sense that this is a robust mathematical concept.  Together with the tight relationship to concepts with known physical applications which one may naturally want to combine, this makes me think physical applications for quantum groupoids may not be so far-fetched after all.

\section*{Ways of ignoring}

Of course, the simpler a physical system is, the more likely we can find an appropriate mathematical abstraction of it, and in practice we do not model the entire system of interest but some simplification of it.    The idea that we can study a complicated system by isolating certain aspects of it to focus on is known as {\em reductionism}. This brings us to what I argue has truly been `unreasonaly effective' in physics.  

We will need to distinguish between two types of reductionism.  At a basic level, reductionism means {\em ignoring} nearly everything in the universe.  
But there are different ways of ignoring: it is one thing to ignore most of the {\em things} in the universe, and quite another to ignore most of the {\em properties} of the things in the universe.  For clarity, I will call the first of these {\bf reductionism} and the second {\bf co-reductionism}.  Both play foundational roles in physics, so it is worth taking some time to understand them in detail.

Mathematically, the difference between  these two types of ignoring corresponds to the difference between {\em sub}-objects and \href{http://ncatlab.org/nlab/show/quotient+object}{{\em quotient} objects}.   A sub-object is essentially what it sounds like: a smaller piece of some other object.  A particular {\em sub}set of set of all people is the set containing only Eugene Wigner, and this can be represented as an arrow from the subset into the whole set, indicating how the part is included in the whole:
\[
\begin{tikzpicture}[xscale=1.1]
\node [draw,rounded corners] {\href{http://en.wikipedia.org/wiki/Eugene_Wigner#mediaviewer/File:Wigner.jpg}{\includegraphics[height=1cm]{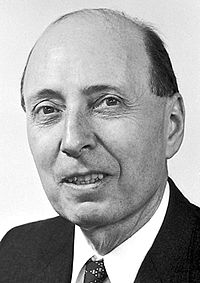}}};
\node at (.65,0) {$\to$};
\node at (2.5,0) [draw,rounded corners] {{\includegraphics[height=1.6cm]{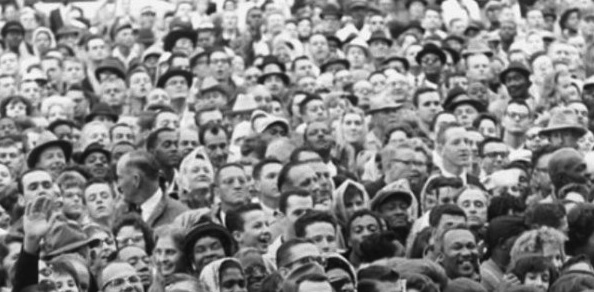}}};
\end{tikzpicture}
\]
We engage in {\em reductionism} when we focus in on a sub-object, for example, to study Wigner as an individual.   Of course, if our goal is to write a biography, we care not just about the Wigner but also his place in the set of all people---that is where the arrow comes in.  However, our main subject is Wigner, and we ignore all other people at least insofar as they do not interact with Wigner.

Quotient objects are less familiar, but are easy to understand by examples.    A particular {\em quotient} of the set of all people is the set of genders.  This can be represented as arrow going from the whole set to the quotient set:    
\[
\begin{tikzpicture}[xscale=1.1]
\node [draw,rounded corners] {$\female\;\mars$};
\node at (-.65,0) {$\to$};
\node at (-2.5,0) [draw,rounded corners] {{\includegraphics[height=1.6cm]{crowd2.jpg}}};
\end{tikzpicture}
\]
The arrow simply represents assignment of the appropriate gender to each person.  Here we are not ignoring {\em any} people, but rather ignoring nearly everything {\em about} all of them.   

Mathematicians have a nice way to think of this: the arrow is thought of as a process in which we ``identify''---declare to be identical---any two source elements that map to the same target element.  In the example, we imagine that Wigner and I, by virtue of being both male, are identified: we {\em  become the same person} according to a vastly oversimplified model where the only distinguishing feature of a person is gender.

It is important to distinguish between `sub-' and `quotient' constructions.  For example, $\female$ and $\male$ do not refer to particular people, so it would be wrong to think of \( \{ \female,\mars\}\)
as a `subset' of the set of people; it is a quotient.  (Confusing this quotient with a subset may be the closest I can come to a mathematically rigorous definition of sexism.)

\section*{Reduction and co-reduction in fundamental physics} 

Every scientific field uses reductionism and co-reductionism, usually a mixture of both, for obvious reasons:  It would be impossible to get off ground in any study without confining our attention to only some of the relevant things, or to some of their relevant properties.  

Often, of necessity, we ignore what cannot be justifiably ignored, and settle for idealized models drastically simpler than the systems we are really interested in.  While this remains a useful tool in all sciences, in physics it appears to be much more.  If physicists are envied by other scientists, it is because the information they ignore has so often turned out to be perfectly justified, at least to within the limits of very precise measurements. 

Reductionism has succeeded in fundamental physics because matter {\em really does} appear to be composed of indivisible and completely identical parts. While the properties of one person imply very little about people in general, the properties of one electron appear to tell us {\em everything} about all other electrons---the are indistinguishable in minutest detail.  Beginning with the atomic theory, and continuing with the discovery of subatomic particles, and finally to the list of about 17 observed particles that physicists now take to be fundamental, we now have built up an astonishingly precise picture of matter---a reductionist's dream, with a finite set of building blocks and neat rules on how to combine them.  

At the other end of the spectrum, what I have called {\em co}-reductionism plays a key role in the other fundamental pillar of modern physics, {\em general relativity}.  In general relativity we ignore almost all properties of matter, considering only its mass.  Remarkably, this again seems to be more than a practical convenience: gravity really does not seem to care about other properties of matter in the least. 

What seems quite miraculous about this whole situation is that gravity, the one force that is so weak that one only notices it when there is a {\em lot} of matter around, is also the one force entirely unconcerned with any distinguishing properties of that matter: it seems to care only about mass and location of that mass.

Quantum gravity, the attempt to combine general relativity with quantum physics, faces a dilemma, since these theories involve orthogonal types of `ignoring.'   Most approaches to the problem can be classified into one of two scenarios.  In one scenario, we try to make gravity more like the rest of physics, for example by studying elementary particles that carry the gravitational interaction (`gravitons') or frameworks to generalize them.  In the other scenario, we attempt to apply principles of quantum physics, which in fact we know entirely from the world of particles, to the relativistic picture of space and time.  

In other words, either we attempt to augment a theory arrived at by reductionist tactics to include a theory arrived at by co-reductionist tactics, or the other way around.  Neither seems obviously bound to work. 
If our luck does run out, it will be a failure of reductionism, and not of mathematics.  In fact, the precise and structural ways of thinking provided by mathematics will be our best hope to surpass this obstacle.

\section*{Factoring a miracle} 

I have argued that what is truly surprising in physics is not that mathematics is effective in physics, but rather that reductionist strategies work so well, for apparently purely physical reasons.  There is also a deeper sense of reductionism lying at the heart of Wigner's argument, and the worldview we are led to if we take it seriously.  

On the surface, Wigner's claim that mathematics is unreasonably
effective in physics seems related to Einstein's famous statement about
the world being amazingly
comprehensible. 
But taken in context, it is clear Einstein does not wonder at the comprehensibility of the world in some metaphysical sense, but rather that the world is comprehensible {\em to us}:

\begin{center}
\begin{tikzpicture}
\node [text width=.6\textwidth,anchor=east,font={\fontfamily{phv}\fontsize{8}{8}\selectfont}]  at (0,0) 
{That the totality of sensory experience is such that it can be organized through thinking \ldots is a fact that we can only marvel at, but which we will never be able to comprehend.  We can say: the eternally incomprehensible thing about the world is its comprehensibility.  };
\node [text width=.3\textwidth,anchor=west,gray,font={\fontfamily{phv}\fontsize{6}{6}\selectfont}] at (0,0) 
{Dass die Gesamtheit
der Sinneserlebnisse so beschaffen ist, dass sie durch das Denken 
\ldots 
geordnet werden k\"onnen, ist eine Tatsache, \"uber die wir nur staunen, die wir aber niemals werden begreifen k\"onnen. Man kann sagen: Das ewig Unbegreifliche an der Welt ist ihre Begreiflichkeit. \cite{Einstein}}; 
\end{tikzpicture}
\end{center}
While it is impressive that one can get so far {\em in principle} with extracting mathematical structure from the physical world, the shocking thing is that we have the mental capacity to do this {\em in practice}.  The real miracle is the complexity of the world {\em relative} to our own intelligence.  This is indeed cause to marvel. It is one thing for the universe to be sensible in some precise way, and quite another for some entity within the universe to {\em make} sense of it to the extent we have.  
	
The desire to `factor out' the human element no doubt stems from reductionist tendencies, where in this case the component we attempt to ignore is {\em ourselves}.  I see no sensible way to do this.

\subsection*{Acknowledgments} 

I am indebted to John Baez, Julian Barbour, David Corfield, and Paul Morris for conversations and insights that have clarified my thinking on this subject, perhaps especially in ways that did not impact this essay itself as much as they might have.

\vfill 
\sf\footnotesize
\noindent 
Department Mathematik, Friedrich-Alexander  Universit\"at Erlangen-N\"urnberg, Cauerstra\ss{}e 11, 91058 Erlangen, Germany. \\[.2em]
{\it Current address:} Concordia University Saint Paul, 
  1282 Concordia Avenue, 
  St.~Paul, Minnesota 55104, USA.
\\[.5em]
derek.wise@fau.de


\newpage 
\begin{thebibliography}{XXXX}
\bibitem[B]{Boehm} G.\ B\"ohm,  Hopf Algebroids, in Handbook of Algebra Vol 6, ed. M. Hazewinkel, Elsevier, 2009, pp. 173-236. 
\bibitem[D-S]{DayStreet} B. Day, R. Street, Monoidal bicategories and Hopf algebroids, {\sl Adv.\ Math.}~{\bf 129} (1997) 99--157.
\bibitem[E]{Einstein2} Albert Einstein, Geometry and Experience, address to the Prussian Academy of Sciences, Berlin, Jan.~1921.  English translation at \\
\url{http://www-history.mcs.st-and.ac.uk/Extras/Einstein_geometry.html}
\bibitem[E2]{Einstein} Albert Einstein, Physik und Realit\"at, {\sl J.\ Franklin Inst.\ }{\bf 221} (1936) 313--347.
\bibitem[I]{Isham} Chris J.\ Isham, Topos methods in the foundations of physics, in {\sl Deep Beauty: Understanding the Quantum World through Mathematical Innovation}, ed. H.\ Halvorson, Cambridge University Press, Cambridge, 2011. 
\bibitem[PCM]{Gowers} {\sl The Princeton Companion to Mathematics}, 
eds. Timothy Gowers, June Barrow-Green, Imre Leader, Princeton University Press, New Jersey, 2008.
\bibitem[GG]{Grattan-Guinness} Ivor Grattan-Guinness, Solving Wigner's mystery: The reasonable (though perhaps limited) effectiveness of mathematics in the natural sciences, {\sl The Mathematical Intelligencer} {\bf 30} (2008) no.~3, 7--17. 
\bibitem[W]{Wigner} Eugene Wigner, The unreasonable effectiveness of mathematics in the natural sciences, {\sl Comm.\ Pure Appl.\ Math.\ }{\bf 13} (1960) 1--14.




\end{thebibliography}
\end{document}